\documentclass[12pt,a4paper]{article}
\usepackage{epsfig,amsmath,amssymb,citesort}

\newcommand{\eqeqref}[1]{Eq.~\eqref{#1}}
\newcommand{\refref}[1]{Ref.~\cite{#1}}

\newcommand{\figref}[1]{Fig.~\ref{#1}}

\newcommand{\beq}{\begin{equation}}
\newcommand{\eeq}{\end{equation}}
\newcommand{\bea}{\begin{eqnarray}}
\newcommand{\beas}{\begin{eqnarray*}}
\newcommand{\beau}[1]{\begin{equation} \label{#1} \begin{array}{rcl}}
\newcommand{\eea}{\end{eqnarray}}
\newcommand{\eeas}{\end{eqnarray*}}
\newcommand{\eeau}{\end{array} \end{equation}}
\newcommand{\bay}{\begin{array}}
\newcommand{\eay}{\end{array}}
\newcommand{\bals}{\begin{align*}}
\newcommand{\eals}{\end{align*}}

\newcommand{\ra}{{\rightarrow}}

\newcommand{\In}[2]{ \left. #1 \right| _{#2} }

\newcommand{\PP}{{\cal P}}

\newcommand{\Nmj}{N^{mj}}
\newcommand{\Nch}{N^{ch}}
\newcommand{\Npart}{N_{part}}

\newcommand{\npp}{n_{p\bar p}}
\newcommand{\spp}{\sigma_{pp}}

\newcommand{\psat}{p_{sat}}

\newcommand{\phobos}{{\sc phobos\ }}

\begin{document}


\begin{titlepage}

\ \vskip1cm
\begin{center}
{\Large \bf Initial conditions and charged multiplicities \\[.2cm] 
in ultra-relativistic heavy-ion collisions} \vspace{1cm} 

        { \bf Alberto Accardi}\footnote{E-mail:{\it accardi@ts.infn.it}} 
	\vspace*{.5cm}  \\
        {\it Dipartimento di Fisica Teorica, Universit\`a di Trieste, \\
        Strada Costiera 11, I-34014 Trieste}  \\
        and \\
	{\it INFN, Sezione di Trieste  \\
    	via Valerio 2, I-34127 Trieste }
  \vspace{.4cm} \\
\end{center}

\vspace{1cm}
\begin{abstract}
At ultra-relativistic energies the minijet production in heavy-ion
collisions becomes sensitive to semi-hard parton rescatterings in the
initial stages of the process. As a result global characteristics of
the event, like the initial minijet density, become rather insensitive   
on the infrared cutoff that separates
hard and soft interactions. This allows to define a nearly
parameter-free {\it saturation cutoff} at which the initial 
conditions may be computed. As an application we study the centrality
dependence of the charged particle multiplicity, which is compared
with present RHIC data and predicted at higher energies. 
\end{abstract}

\begin{footnotesize}
\centerline{PACS: 11.80.La, 24.85.+p, 25.75.-q}
\end{footnotesize}

\end{titlepage}

\newpage 

\setcounter{page}{2}
\setcounter{footnote}{0}


\section{Introduction}
\setcounter{equation}{0}

In heavy-ion collisions the partonic degrees of freedom of the two
interacting nuclei become more and more important as the center of
mass energy of the collision increases. At some point the main
particle production mechanism in the initial stage becomes the
liberation from the nuclear wave functions of a great number of
partons, also called {\it minijet plasma}. 
At ultra-relativistic energies the partonic density of the two nuclei
is so high that perturbative methods on one hand
\cite{inpert,KLL87,EKRT00,EKRT2,WG01,KN01} 
and semi-classical non-perturbative methods on the other \cite{innonpert}
become applicable to the computation of the initial conditions of the
minijet plasma. It's successive evolution will
possibly lead to thermalization of the system and to the transition to
the quark-gluon plasma phase, whose formation and characteristics
depend crucially on such initial conditions. Though the latter are not
directly accessible experimentally, they can be related to final state
observables, like the charged particle multiplicity and transverse
energy, allowing a test of the proposed theoretical models. 

We can divide in general the models in three classes:
{\it i)} two-component models \cite{WG01,KN01},
in which particle production is assumed to be decomposable into the sum
of a soft and a hard part according to some cutoff $p_0$;
{\it ii)} saturation models \cite{inpert,EKRT00,EKRT2,innonpert}, which
exploit the high parton densities involved in the process;
{\it iii)} ``others'', like the Dual Parton Model
\cite{CS01} and hydrodynamic models \cite{KHHET01}. 
To distinguish between them, it has been
proposed in \cite{WG01} to study the centrality dependence of the
charged particle multiplicity, since this allows to disentangle to
some degree the dynamical and the geometrical effects. For a review of
the results of the above models on the charged multiplicity see
\cite{Esk01}. 

At very high energies the target parton densities experienced by 
projectile partons are so high that the probability for them to have more
than one semi-hard scattering may become non negligible already at the
BNL Relativistic Heavy-Ion Collider (RHIC). At such regimes the usual
perturbative computation \cite{KLL87}, obtained by eikonalization of the
minijet cross-section, may become inadequate. Indeed, it takes into
account only disconnected two-parton interactions located at different
points in transverse space but neglects the rescatterings.
With the help of a few simplifying hypotheses semi-hard parton rescatterings
have been included in the interaction mechanism in \cite{CT90,CT91+},
and lead to sizeable effects already at RHIC energies
\cite{AT01a,Acc01,AT01b}. Based on these results, in this paper we propose a
new saturation mechanism for semi-hard minijet production and use it
in a two-component model to compute charged particle
multiplicities at RHIC and at the CERN Large Hadron Collider (LHC).


\section{Initial conditions and saturation}
\setcounter{equation}{0}

When rescatterings are included in the interaction of two nuclei of atomic numbers A and B, the average number of A nucleus minijets at fixed impact parameter $b$ is given by \cite{CT90}:
\begin{align}
     \Nmj_A(b)=\int d^2r dx \Gamma_A(x,b-r) 
	\Bigl[1-e^{-k\int dx' \sigma_H(xx') \Gamma_B(x',r)}\Bigr] \ ,
    \label{NmjA}
\end{align}
and the average minijet initial multiplicity is obtained by summing the
analogous contribution from the B nucleus, $\Nmj = \Nmj_A + \Nmj_B$.
For simplicity we omit the flavour indices and consider only
gluon-gluon interactions in our formulae, the inclusion of quarks
being straightforward. In the numerical computations both the gluons
and the quarks have been included. 
In \eqeqref{NmjA}, $\Gamma_A=\tau_A(r)G(x)$ is the nuclear parton
distribution function of the A nucleus, $\tau_A(r)$ is
its nuclear thickness function, normalized to A, evaluated at a transverse
coordinate $r$ relative to the center of the nucleus and $G(x)$ is the
parton distribution function of a proton at a given fractional momentum
$x$. For simplicity we omit the flavour indices. $\sigma_H$ is the pQCD gluon-gluon cross-section at leading order in the high energy limit,
\begin{align*}
	\sigma_H(xx') = \frac92 \pi \alpha_s^2 \frac{1}{p_0^2}
		\left( 1 - \frac{4p_0^2}{xx's} \right)
		\theta(xx's - 4p_0^2)\theta(1-x)\theta(1-x') \ ,
\end{align*}
where we included all the kinematic limits and $p_0$ is the cut-off
that discriminates between soft and semi-hard interactions. We also included
explicitly the $k$-factor, $k$, to take into account higher order
corrections. Both the cross section and the parton distributions
depend on a scale $Q=p_0$, which we take equal to the cutoff.
In the numerical computations we will set $k=2$ and use Woods-Saxon
thickness function and GRV98LO parton distribution functions \cite{GRV98}.

\eqeqref{NmjA} may be interpreted as the integral of the average density
of projectile partons (at a given $x$ and $r$) times the probability of
having at least one semi-hard scattering against the target. The
exponent in \eqeqref{NmjA} may be interpreted as the opacity of the
target nucleus, being proportional to the total transverse area
occupied by its partons at the resolution scale $p_0$.
Two interesting limiting cases may be studied. At high values of $p_0$
the target has a small opacity and is seen by the incoming partons as
a rather dilute system. As a consequence $\Nmj \approx 2 \int d^2r dx dx'
\Gamma_A(x,b-r)\sigma_H(xx')\Gamma_B(x')$, and we recover the usual
perturbative result \cite{KLL87}. On the other hand, at low values of 
$p_0$ the target opacity increases: the target is becoming black to the
projectile partons. As a consequence, the probability of scattering 
at least once becomes so high that nearly every projectile parton
scatters and the minijet multiplicity reaches a limiting value
instead of diverging as it happens in the Eikonal computation. 

\begin{figure}[t]
\begin{center}
\vskip-.7cm 
\epsfig{figure=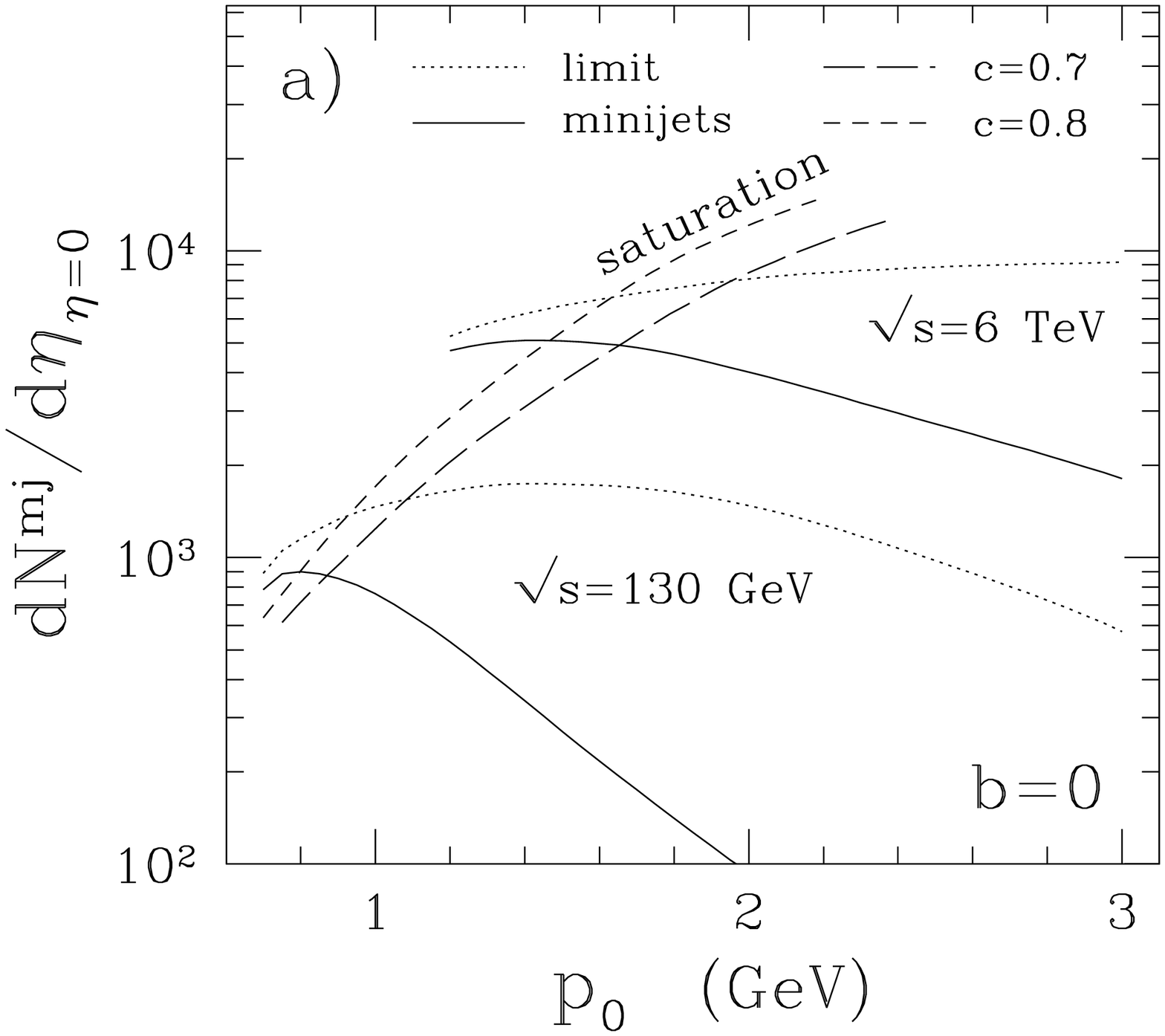,width=5cm}
\epsfig{figure=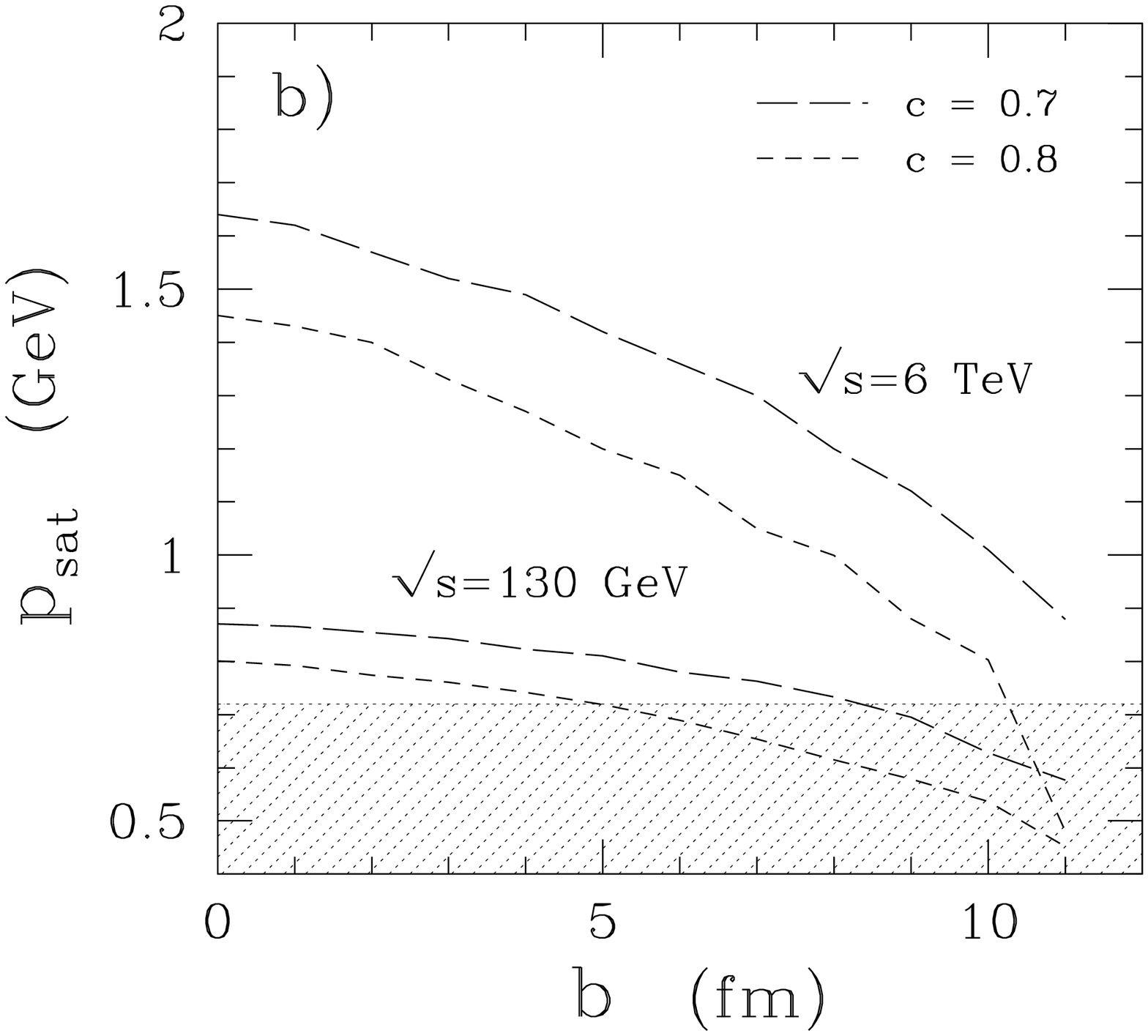,width=5cm}
\caption{\footnotesize
$a)$ The minijet multiplicity, $\Nmj$ (solid line), and its limiting
value, $\Nmj_{lim}$ (dotted line), in a central Au-Au collision as a
function of the cutoff $p_0$ at RHIC and LHC energies. The dashed
lines are the saturated minijet multiplicities, 
$\Nmj_{sat}$, with a saturation parameter $c=0.7$ (long dashes) and
$c=0.8$ (short dashes). The intercept of the dashed lines with the
dotted lines determines the saturation cutoff. 
$b)$ The saturation cutoff as a function of the impact parameter at
RHIC and LHC energies. The shaded area is the region where we estimate
that the saturation criteria ceases to be valid.  
  \label{fig:psat} } 
\end{center}
\end{figure}

In the regime where the target is almost black the semi-hard
interactions are extracting 
from the projectile nucleus wave-function all its partons, and even if
we use a lower cutoff no more partons are there to be extracted. for
this reason the minijet multiplicity tends to saturate \cite{AT01a}, see also
\figref{fig:psat}a. We call {\it saturation cutoff} the value of $p_0$
at which this happens, and will denote it as $\psat$.   Of course the
validity of this picture is limited to the kinematic regions where the
saturation cutoff is in the perturbative range, 
$\psat \gg \Lambda_{QCD}$. 
To give a quantitative definition of the saturation cutoff we start by
considering a central collision of two equal nuclei. We define the
{\it upper bound} for the minijet multiplicity as
\begin{align}
     \Nmj_{Alim}(b=0)=\lim_{k\ra\infty} \Nmj_{A} 
	= \int_{4p_0^2/s \leq x \leq 1} \hspace*{-1.3cm}
	d^2r dx \Gamma_A(x,r) \ .
 \label{NAlim}
\end{align}
Taking a very large $k$-factor corresponds, indeed, to the limit in
which the target becomes completely black and the semi-hard
interactions are effective in extracting all the partons from the
projectile nucleus. The limiting procedure is needed in order to keep
track of the kinematic limits. As it is easy to see, 
$\Nmj \underset{p_0 \ra 0}{\sim} \Nmj_{lim}$, 
therefore we can define the saturation cutoff as
the value of $p_0$ such that the minijet multiplicity becomes a
substantial fraction of its limiting value:
\begin{align}  
	\Nmj(p_0=\psat) = c \, \Nmj_{lim}(p_0=\psat) \ ,
 \label{psat}
\end{align}
where the {\it saturation parameter}, $c$, is a positive number
smaller than one. Notice that $\psat = \psat(\sqrt s,c)$ is a function
also of the energy of the collision. From our discussion it is obvious that $c$ must be close to one to let
$\psat$ lie in the region where $\Nmj$ is saturating. However, to stay
in the perturbative regime we cannot choose it too close to one since
$\psat \ra 0$ as $c \ra 1$. Finally, we define the {\it saturated
minijet multiplicity} as the average multiplicity evaluated at the saturation
cutoff: 
\begin{align}
	\Nmj_{sat} = \Nmj_{sat}(\sqrt s,c) = \Nmj(p_0=\psat) \ .
 \label{mjsat}
\end{align}
In our approach this number represents also the multiplicity of
partons produced in the early stage of the heavy ion collision.

In Fig.~\ref{fig:psat}a we
show the minijet multiplicity and its limiting value as a function of
the cutoff $p_0$ at RHIC and LHC energies. The rapidity density at
$\eta=0$ is computed by integrating Eqs.~\eqref{NmjA} and
\eqref{NAlim} over a pseudo-rapidity interval $|\eta| \leq 1$, where
we approximated $\eta \approx \log\left( x\sqrt s /p_0 \right)$, and by
dividing the result by a factor two. The dashed lines represent
the saturated initial conditions computed with $c=0.7$ and $c=0.8$. 
We can see that at a given energy $\Nmj_{sat}$, which is obtained
as the intercept of the solid and dashed lines, is nearly
independent of the saturation parameter as long as the latter is close
enough to one. Indeed, both at RHIC and LHC energy we obtain
approximately a 3\% increase in the saturated multiplicity going 
from $c=0.7$ and $c=0.8$. Therefore, whereas $c$ is an arbitrary
parameter its actual 
choice doesn't affect strongly the determination of the initial
conditions. The dependence of $\psat$ on $\sqrt s$ was studied in
\cite{Acc01} where it is shown that for central collisions the
saturation criterion is applicable from RHIC energies on. 

Unless we use nuclear thickness functions with sharp edges, like the
hard-sphere distributions, by applying blindly the saturation
criteria to non central collisions we would obtain an impact
parameter independent bound on the minijet multiplicity. Indeed we
would have $\Nmj_{lim}(b) = \int d^2r dx \Gamma_A(x,b-r) = \int d^2r
dx \Gamma_A(x,r)$. In this way, by requiring saturation as in \eqeqref{psat}
 we would be asking the semi-hard interactions to extract all the
partons from the projectile nucleus even in a very peripheral region,
which is clearly unphysical. A simple way to implement the collision geometry in the saturation
criterion is to cut by hand the thickness functions 
outside a given radius $R_c$ of the order of the nuclear radius. 
However, the minijet multiplicity as a function of the
centrality of the collision turns out to depend too strongly on the
choice of $R_c$ except at very high centrality or very high energies 
\cite{Acc01}. 

To find a less arbitrary way of implementing the collision geometry we
look at the Glauber model computation of the average number of nucleons
which participate in the collision:
\begin{align}
	\Npart(b) = \int d^2r \, \tau_A(b-r) \PP_B(r) 
		+ A \leftrightarrow B \ ,
 \label{Npart}
\end{align}
where $\PP_B(r)  = 1 - \left[ 1 - \spp(s){\tau_B(r)}/{B} \right]^B$
and $\spp$ is the inelastic $pp$ cross-section, which we take from
\cite{PDB}. At $\sqrt s=130,200,6000$ GeV we have $\spp=39,42,75$
mbarn, respectively. $\PP_B$ is the probability that a projectile nucleus at a
given transverse coordinate $r$ has at least one inelastic interaction
with the target nucleons. Then, we may require the saturation only for
the fraction of projectile partons that belong to a participating
nucleon, and define an {\it effective nuclear distribution function} 
\begin{align*}
	\overline\Gamma_{AB}(x,b,r) = \Gamma_A(x,b-r)\PP_B(r) \ .
\end{align*}
Correspondingly, we have an effective minijet multiplicity, 
$\overline N^{mj}_A(b) = \int d^2r dx \overline\Gamma_{AB}(x,b,r) 
\big[ 1 - \exp\big(-\int dx'\sigma_H(xx')\Gamma_b(x')\big) \big] $, and
an effective upper limit, $\overline N^{mj}_{Alim}(b) = \int d^2r dx
\overline\Gamma_{AB}(x,b,r)$, which is no more $b$-independent.
Then, the saturation criterion generalized to arbitrary impact
parameter becomes:
\begin{align}  
	\overline N^{mj}(p_0=\psat) = 
		c \, \overline N^{mj}_{lim}(p_0=\psat) \ .
 \label{psatb}
\end{align}
Finally, having determined $\psat$ in this way we use it in
\eqeqref{mjsat} to compute the average initial parton multiplicity. 

In Fig.~\ref{fig:psat}b we show the saturation cutoff as a function of
the impact parameter at RHIC and LHC energies for different saturation
parameters $c$. The horizontal line show the limit of approximately
0.7 GeV whose intersection with $\psat(b)$ sets the limit of validity
of the present approach, as will be discussed in the next section.

Notice that the saturation cutoff, and consequently the initial
conditions, are practically determined by the choice of the
parton distribution functions. As explained above and in \cite{AT01a}
the initial conditions are nearly independent of the remaining free
parameters, namely the saturation parameter c and the $k$-factor.

\begin{figure}[t]
\begin{center}
\vskip-.7cm 
\epsfig{figure=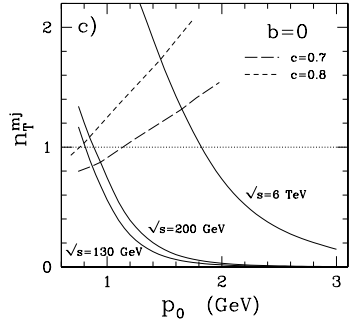,width=5cm}
\caption{\footnotesize
The minijet average occupation number in the transverse area (solid
lines): $n_T^{mj} = A_T^{mj}/A_T^{Au}$, where  
$A_T^{mj}=\Nmj(p_0) \times \pi/p_0^2$ and $A_T^{Au}$ is the transverse
area of a gold nucleus. The dashed lines show the average occupation number
of the saturated minijets as a function of the saturation momentum.
When $n_T^{mj} \gtrsim 1$ the minijets begin to overlap transversely.
  \label{fig:trarea} } 
\end{center}
\end{figure}

In the proposed mechanism saturation is reached when there are no more
partons that semi-hard interactions can extract from the nuclear wave
functions. In this sense this mechanism is a saturation of the minijet
production 
and is intermediate between initial and final state saturation. In 
initial state saturation \cite{inpert} (see also
\cite{innonpert,KN01}) 
the parton density inside the incoming nuclei saturate due to a
compensation between parton splitting and parton 
fusion processes in the DGLAP evolution, which
induces a corresponding saturation in the minijet multiplicity. On the
contrary, in the final state mechanism \cite{EKRT00,EKRT2} the
saturation is assumed to be caused by the high 
density of produced minijets, which screens softer parton production
due to parton fusion processes in the final state. In particular these
final state interactions are assumed to set in when the transverse
area occupied by the minijets becomes comparable to the nuclear
overlap area. 
Both processes may, therefore, complement our saturation mechanism
since the former modify the input parton distribution
functions and the latter deals with a later stage
process. However, as we can see in Fig.~\ref{fig:trarea}, at $\sqrt
s=130$ GeV and $\sqrt s=200$ GeV the saturated minijets 
fill the transverse area only partially. Therefore final state
saturation effects should not alter significantly our computations at
RHIC energy,  but may play some role at LHC.


\section{Charged particle multiplicity}
\setcounter{equation}{0}

We want to apply the saturation criterion for the semi-hard parton
production in the initial stage of the collision to the computation of
the charged particle multiplicity. Thanks to the self-shadowing
property of the semi-hard interactions
\cite{selfshad,AT01b}, even if in \eqeqref{NmjA} only the semi-hard
cross-section, $\sigma_H$, appears, we are actually taking into
account all the partons that had {\it at least} one semi-hard
scattering, while their other scatterings may be semi-hard or soft
with no restrictions. Therefore we are missing only the
purely soft part of the production mechanism. This leads us to adopt a
two-component model in which the charged particle multiplicity
is written as the sum of a soft and a semi-hard part: 
${d\Nch}/{d\eta}(b) =
{d\Nch_{soft}}/{d\eta}(b)+{d\Nch_{s.h.}}/{d\eta}(b)$.  
The soft part is assumed to scale with the
number of participants, \eqeqref{Npart}, so that \cite{KN01}
\begin{align}
	\frac{d\Nch_{soft}}{d\eta}(b) = x \npp(s)
			\frac{\Npart(b)}{2} \ .
  \label{nch2}
\end{align}
Here $\npp(s)$ the pseudo-rapidity density of charged particles
produced  at $\eta=0$ in $p \bar p$ collision at a given 
c.m. energy $\sqrt s$. We use the fit 
\cite{CDF90}, $\npp(s) = 2.5 - 0.25 \log(s) + 0.023 \log^2(s)$.
The coefficient $x=x(s)$ is a parameter that allows to adjust
the relative weight of soft and semi-hard interactions and will be
determined from the experimental data. 
Further, we assume the semi-hard part to be completely computable from the
saturation criterion for minijet production described in the last section.
To convert the minijet multiplicity to charged particle multiplicity,
we further assume isentropic expansion of the initially produced
minijet plasma and parton-hadron duality, so that
\begin{align}
	\frac{d\Nch_{s.h.}}{d\eta}(b) = 0.9 \times \frac{2}{3} \times
		 \frac{d\Nmj_{sat}}{d\eta}(b)  \ ,
  \label{nch3}
\end{align}
where the factor $0.9$ is due to the different number of degrees of
freedom of the system in the minijet-plasma phase and in the hadronic
phase \cite{EKRT00}. To mark out the contribution of the hard part it
is customary to divide the charged multiplicity by the number of
participant pairs, so that the observables we are interested in are:
\begin{align}
	\frac{1}{\Npart(b)/2} \frac{d\Nch}{d\eta}(b) 
		= x \, n_{pp}(s)
		+ \frac{1}{\Npart(b)/2} \frac{d\Nch_{s.h.}}{d\eta}(b) 
 \label{nch1}
\end{align}  
and the fraction of semi-hard interactions, $F_{s.h.} =  
\frac{d\Nch_{s.h.}}{d\eta} / \frac{d\Nch}{d\eta}$.

To make a comparison with experimental data we have first to relate the
observables appearing in \eqeqref{nch1}, which are functions of the
impact parameter, to the experimental ones, which are obtained as
averages over centrality classes of events
\cite{PHOBOS00,PHOBOS01,PHENIX01}. Following \cite{KN01,KLNS97}, to
which we refer for the details, we do this by studying the minimum
bias multiplicity distribution of charged particles and by dividing
the events in suitable subsets over which the average is performed. 
The next step is to extract the parameter $x$ in \eqeqref{nch2} 
by comparing the computation for the 3\% most central
events and the \phobos data at $\sqrt s = 130$ GeV from
\refref{PHOBOS01}. This value is then 
used to make predictions at higher energy.

\begin{figure}[t]
\begin{center}
\vskip-.7cm 
\epsfig{figure=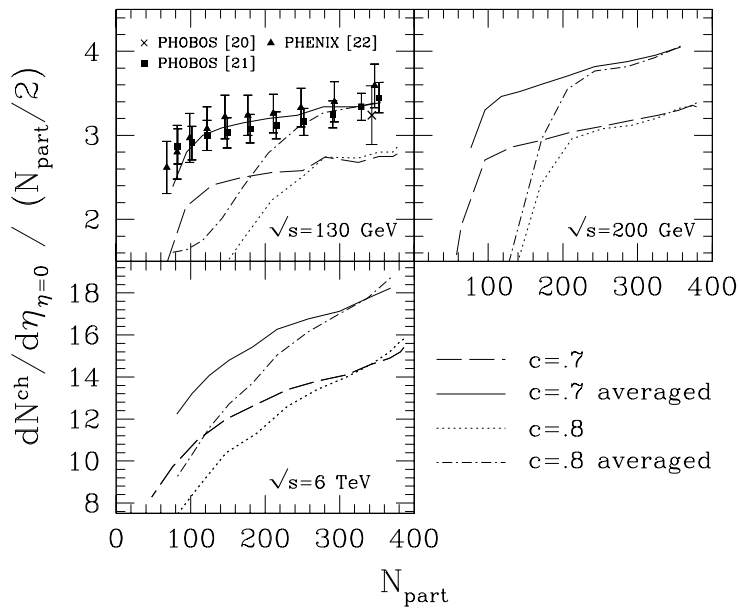,width=10cm}
\vskip-.5cm 
\caption{\footnotesize
Charged particle multiplicity per participant pair, \eqeqref{nch1}, 
as a function of the number of participants at different center of
mass energies and saturation parameter $c=0.7$ and $c=0.8$.
In each panel the lower pair of curves (dashed and dotted lines)
represent the semi-hard contribution. The upper pair of curves (solid
and dot-dashed lines)
are obtained by averaging the semi-hard contribution and by adding the
soft part. At $\sqrt s = 130$ GeV the parameter $x$ is
extracted from the 3\% most 
central \phobos events at $\sqrt s = 130$ GeV \cite{PHOBOS01}. At higher
energies the upper curves may be considered an upper bound, while the
lower ones give a lower bound, see text.
  \label{fig:chmult} } 
\end{center}
\end{figure}

In \figref{fig:chmult} we show both the results for the semi-hard part
before the averaging over the centrality classes, and the
results obtained after the averaging and the inclusion of the soft
part.  For each curve the result obtained by setting $c=0.7$ and
$c=0.8$ in \eqeqref{psatb} is shown.

At $\sqrt s = 130$ GeV we find $x=0.445$ and $x=0.453$, for a
saturation parameter $c=0.7$ and $c=0.8$, respectively. 
These values of $x$ correspond to a fraction of
semi-hard interactions $F_{s.h.}=0.805$ and $F_{s.h.}=0.817$,
respectively, and show a good stability with respect to $c$. 
The relatively large value of $F_{s.h.}$ with
respect to the common expectation of nearly a half and to the value of 0.37
extracted from \phobos data in \refref{KN01} is due to the fact that we
considered as belonging to the non-soft part of the observable also
a {\it semi}-hard region $0.7$ GeV $\lesssim p_0 \lesssim 2$ GeV. Note
that we can push our perturbative computations to such low values of
the cutoff because inclusion of parton rescatterings results in a
rather small sensitivity of global observables on $p_0$ in that region
\cite{AT01a,Acc01}.\\ 
The two curves start with a moderate slope at high centrality and at
some point they decrease very fast. This happens when the
corresponding saturation cutoff becomes smaller than 0.7 GeV,
approximately. The reason for this behaviour is that the distribution
functions are fitted just down to a scale $Q \approx 0.9$ GeV and they
are numerically extrapolated at lower scales. Below 0.7 GeV the
extrapolation gives an unnaturally fast decrease of the parton densities,
which results in the rapid fall of the minijet production. Then, we
define the region of validity of our computations as the one such that
$p_{sat} \gtrsim 0.7$ GeV, or in other words the one to the
right of the knee in the charged multiplicity.  \\
The value of $p_{sat}$ at fixed centrality decreases when the
saturation parameter $c$ increases (see \figref{fig:psat}), therefore
the curve with $c=0.8$ 
is reliable for a smaller range of centrality than the curve with
$c=0.7$. They agree, however, in the common region of validity (showing
a slight tendency to increase their slope when increasing
$c$), and after the experimental averaging and the fit
to the most central data point, both describe well the experimental
data.

At $\sqrt s = 200$ GeV we don't have any data to normalize the
multiplicities to. However, the fraction of semi-hard to soft
interactions is expected to grow with the energy of the collision, and
we can use the value of $F_{s.h.}$ determined at $\sqrt s = 130$ GeV
to obtain an approximate upper bound for the charged multiplicities:
for $\sqrt s \geq 130$
\begin{align}
	x\npp(s) \leq \frac{
		\In{1-F_{s.h.}}{b=0,\sqrt s= 130{\rm\ GeV}} 
		}{
		\In{F_{s.h.}}{b=0,\sqrt s= 130{\rm\ GeV}} 
		} 
		\frac{d\Nch_{s.h.}/{d\eta}}{\Npart/2}(b=0,s) \ .
\end{align}
The curves for
the two values of $c$ agree over a wider range of neutralities. This
is to be expected since the saturation cutoff at fixed centrality
grows with the center of mass energy, and goes below the critical value
of $0.7$ GeV at smaller centrality. Notice also that the slope of the
curves has increased.

At LHC energy, $\sqrt s = 6$ TeV, the particle production is generally
believed to be almost completely semi-hard. Therefore we expect that the data
will be close to the averaged semi-hard multiplicity without any
normalization (which is very similar to the lower curve plotted in 
\figref{fig:chmult}). Though the saturation criterion is applicable
over all the centrality range considered (see Fig.~\ref{fig:psat}b)
the slope of the curves is rather sensitive to the saturation
parameter, resulting in a larger theoretical uncertainty. We expect
that a better treatment of the scale $Q$ and of the pseudo-rapidity,
which are taken to depend simply on the cutoff $p_0$, could solve at
least partially this problem. However, the average slope has increased
confirming the trend observed at lower energies.


\section{Conclusions}
\setcounter{equation}{0}

The inclusion of semi-hard parton rescatterings in the interaction
dynamics of heavy-ion collisions at very high energy allows a reliable
computation of the initial conditions, like the minijet multiplicity,
and the introduction of a nearly parameter-free saturation criterion
to determine the infrared cutoff to be used in the perturbative
computations. The proposed saturation mechanism is intermediate
between the initial and final state ones in that it deals with the
saturation in the production of minijets. 
 
We tested our approach against RHIC data on the centrality dependence
of charged multiplicities by using a two-component model in which the
semi-hard part is assumed to be completely given by the proposed
saturation criterion. At $\sqrt s=130$ GeV we find a good
agreement with the data, which allows us to extrapolate the results
at the highest RHIC energy of $\sqrt s=200$ GeV and at LHC energy,
$\sqrt s=6$ TeV, by putting upper and lower bounds on the charged
multiplicities per participant pairs as a function of the number of
participants and by predicting their slope.

\vskip.25in 
{\bf Acknowledgments} 
\vskip.15in 

I am very grateful to D.~Treleani and P.~Huovinen for their
stimulating comments and a critical reading of the
manuscript. I would like 
also to thank R.~Arnaldi, M.~Gyulassy,
D.~Kharzeev, H.~J.~Pirner, A.~Polleri, C.~Salgado
U.~Wiedemann and F.~Yuan for many useful discussions. 
This work was partially supported by the
Italian Ministry of the University and of the Scientific and Technological
Research {\sc (MURST)} by Grant No. {\sc COFIN99}.


\end{document}